\documentclass[pra,twocolumn,a4paper]{revtex4}

\usepackage{graphicx}
\usepackage{dcolumn}
\usepackage{amsmath}
\usepackage{amssymb}
\usepackage{subeqnarray}
\usepackage{bm}
\usepackage[usenames]{color}

\newcommand\ket[1]{\mid\hspace{-0.1cm} #1\rangle}
\newcommand\bra[1]{\langle#1\mid}
\newcommand\bk{{\bf k}}
\newcommand\bq{{\bf q}}
\newcommand\bx{{\bf x}}
\newcommand\bkcm{{\bf K}}
\newcommand\veck{\vec k}
\newcommand\vecq{\vec q}
\newcommand\vecr{\vec r}

\begin{document}

\title{Generation of hyper-entangled photon pairs in coupled microcavities}
\author{S. Portolan$^1$, L. Einkemmer$^2$, Z. V\"or\"os$^2$, G. Weihs$^2$, P. Rabl$^1$}
\affiliation{$^1$Institute of Atomic and Subatomic Physics, TU Wien, Stadionalle 2, 1020 Wien, Austria}
\affiliation{$^2$Institut f\"ur Experimentalphysik, Universit\"at Innsbruck, Technikerstr. 25, 6020 Innsbruck, Austria}

\begin{abstract}
We propose and theoretically analyze a new scheme for generating hyper-entangled photon pairs in a system of polaritons in coupled planar microcavities. Starting from a microscopic model, we evaluate the relevant parametric scattering processes and numerically simulate the phonon-induced noise background under continuous-wave excitation. Our results show that, compared to other polariton entanglement proposals, our scheme enables the generation of photon pairs that are entangled in both path and polarization degrees of freedom, and simultaneously leads to a strong reduction of the photoluminesence noise background. This can significantly improve the fidelity of the entangled photon pairs under realistic experimental conditions.
\end{abstract}




\maketitle

\newpage

Entanglement is considered the primary resource for quantum information processing schemes and in the optical domain the practicability of implementing large-scale photonic quantum computers or long distance quantum communication protocols relies crucially on efficient sources for entangled photon pairs (EPPs). Conventionally, EPPs are produced by parametric down conversion in nonlinear crystals \cite{Mandel} or four-wave mixing in photonic crystal fibers~\cite{CohenPRL2009}. Solid state systems like exciton polaritons in microcavities~\cite{Rassegna Polaritons} offer an intriguing alternative with the prospect of building highly efficient EPP sources on a miniaturized scale~\cite{SavvidisPRL2000,SabaNature2001}. 
Being half exciton, half photon, polaritons benefit from strong Coulomb interactions, while they can easily be converted into propagating optical qubits for long-distance entanglement distribution. Over the past years polariton-polariton interactions in microcavities have been the subject of intensive research~\cite{StevensonPRL2000,ErlandPRL2001,DiederichsNature2006,CiutiSST2003,SavastaPRBPRL}. However, the predicted quantum properties of generated polariton pairs~\cite{SavastaSSC99,CiutiPRB2004,PortolanEPL09} have remained quite elusive and no direct experimental evidence for entanglement detection has been demonstrated so far, mainly due to background noise caused by phonon-induced photoluminescence~\cite{LangbeinPRB2004}.

\begin{figure}[b]
\begin{center}
\includegraphics [width=0.45\textwidth]{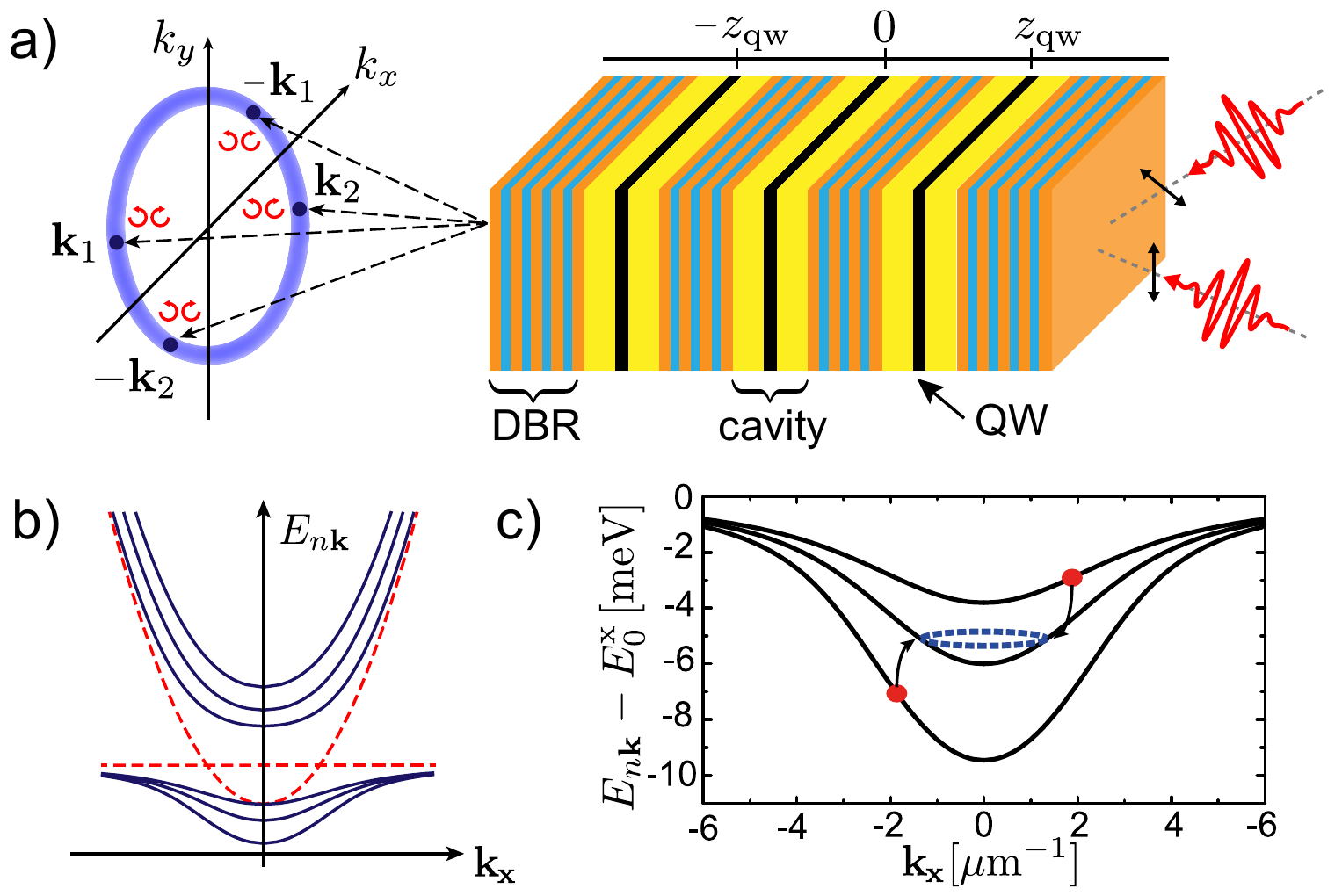}
\caption{(Color online). 
a) Sketch of the three coupled cavities, which are separated by two distributed Bragg reflectors (DBR) and in each cavity spacer a quantum well (QW) is inserted~\cite{DiederichsNature2006}.
b) Energy dispersion curves of the 6 polariton branches (solid lines); the dashed lines show the energies of the bare exciton and the lowest photonic mode. c) Actual lower polariton branches for the sample under consideration. 
(see text). The red circles show the chosen pump wavevectors $(\bk_p = -\bk'_p, |\bk_p| = 1.8 \,\mu{\rm m}^{-1})$ which, by mixed parametric scattering, generate a circle of energy-degenerate hyper-entangled signal-idler pairs (dashed blue line).}\label{Coupled_cavities}
\end{center}
\end{figure}

In this work we investigate the potential of coupled microcavity structures 
as an efficient platform for semiconductor quantum technologies and apply this concept to the design of bright sources of entangled and hyper-entangled photon pairs.
In particular, we consider a system of three planar microcavities~\cite{DiederichsNature2006}, which are coupled via two shared Bragg mirrors as illustrated in Fig.~\ref{Coupled_cavities} a). In this setup the splitting of the upper and lower polariton dispersion curves into three well-separated sub-branches provides an additional flexibility for engineering parametric inter-branch scattering processes~\cite{CiutiPRB2004}, which leads to qualitative new features and an improved performance of the EPP creation. 
By pumping two polariton modes in the first and third branch, the phase matching condition for parametric scattering is fulfilled on an energetically degenerate circle of momentum states in the middle branch as shown in Fig.~\ref{Coupled_cavities} c). By choosing different pumping configurations, photon pairs with polarization entanglement or pairs with entanglement in both the path \emph{and} polarization degrees of freedom can be generated. Such \emph{hyper-entangled} photons~\cite{KwiatJMO1997,BarbieriPRA2005,BarreiroPRL2005} provide a valuable resource for super-dense coding protocols \cite{WalbornPRA2003,BarreiroNatPhys2008} or quantum key distribution \cite{WangJOSAB2009}. Moreover, 
the energy separation between the different branches creates a bottleneck for phonon-induced polariton scattering and under realistic conditions suppresses the background photoluminesce by several orders of magnitude. 
Our analysis of the resulting entanglement fidelity 
shows that the proposed multi-cavity setting offers a competitive option for future solid state based EPP sources and integrated quantum technologies.

{\it Model.---}  
In the setup shown in Fig.~\ref{Coupled_cavities} a) the photons in each cavity are coupled to electronic interband excitations (excitons) of a semiconductor quantum well (QW) and form new quasi-particles -- so-called polaritons -- under strong coupling conditions. We assume that both photons and excitons are restricted to a single mode function along the confined direction (z-axis) and we denote by $E^c_{n \bk}$ ($E^\text{x}_{n \bk}$) the energy dispersion of photons (excitons) with transverse momentum $\bk$ and cavity index $n$. The Hamiltonian for the whole system is  
 \begin{align}\label{H}
\hat H = & \sum_{n,\bk} E^c_{n \bk} \hat a^\dag_{n \bk} \hat a_{n \bk} + E^\text{x}_{n \bk} \hat B^\dag_{n \bk} \hat B_{n \bk} + V_n \left( \hat a^\dag_{n \bk} \hat B_{n \bk} + {\rm H.c.}\right)  \nonumber \\
 & - \sum_{n,\bk}  J_{n,n+1}\left( \hat a^\dag_{n \bk} \hat a_{n+1 \bk} + {\rm H.c.} \right) + \hat H_{\rm C},
 \end{align}
where $\hat a_{n \bk}$ and $\hat a^\dag_{n \bk}$ are bosonic annihilation and creation operators for photons in cavity $n$ and transverse momentum $\bk$. $\hat B_{n \bk}$ and $\hat B^\dag_{n \bk}$ are the corresponding operators for excitons, which  
can be treated as effective  bosons under low excitation conditions \cite{suppinfo}.  The so-called Rabi splitting $V_n$ is the dipole coupling strength between photons and excitons within cavity $n$ and $J_{n,n+1}$ is the optical mode coupling between neighboring cavities.  Finally, $\hat H_{\rm C}$ accounts for the Coulomb interaction between excitons, which will give rise to the effective $\chi^{(3)}$ polariton-polariton non-linearity discussed below.

In the following we write $\hat H =\hat H_0+\hat H_{\rm int}$, where apart from the Coulomb term, $\hat H_{\rm int}$ accounts for additional effective interactions due  to the non-bosonic character of the excitons~\cite{CiutiSST2003,PortolanPRB2008_DCTS}. The linear part, $\hat H_0$, can be diagonalized by a generalized Hopfield transformation $\hat H_0=\sum_{i\bk} E_{i\bk}   \hat P^\dag_{i \bk} \hat P_{i \bk}$~\cite{BassaniNuovoCimento1986}, and expressed in terms of a set of polaritonic quasi-particle operators 
 \begin{equation}
 \hat P_{i \bk} = \sum_n X^n_{i \bk} \hat B_{n \bk}+ C^n_{i \bk} \hat a_{n \bk},
 \end{equation}
where $X^n_{i \bk}$ and $C^n_{i \bk}$ are the exciton and photon components of the $i$-th polariton branch. For a single cavity the index $i$ denotes the familiar lower and upper polariton branches, which, for zero detuning, are split by $\sim 2 V$ around $\bk=0$ \cite{Coupled_Oscillators_Model}. In coupled structures, the optical mode coupling induces a further splitting  
of $\sim J$.  Typical dispersion curves for three coupled cavities are shown in Fig.~\ref{Coupled_cavities} b), where  identical Rabi splittings $V=6$ meV and mirror couplings $J=4$ meV have been assumed.

{\it Photon pair creation.---} The nonlinear optical response of our system is governed by the strong Coulomb interaction between excitons, $\hat H_{\rm int} \sim \hat B^\dag \hat B^\dag \hat B \hat B$, which results in an equivalent $\chi^{(3)}$-type non-linearity between polaritons.  In the following we are focusing on the lower set of branches $i=1,2,3$, which have a smaller exciton-exciton dephasing \cite{SavastaPRBPRL,PortolanEPL09}. We assume that two polariton modes with wavevectors $\bk_p$ and $\bk_p^{\prime}$ and energies $E_{\bk_p}$ and $E_{\bk_p^{\prime}}$, respectively, are strongly pumped by external lasers and we linearize the interaction around the classical mean value of these two pumped modes. We obtain a parametric scattering process analogous to four-wave-mixing in nonlinear optical crystals, 
\begin{equation}\label{chi3}
\hat H_{\chi^{(3)} }= \sum_{k_s,k_i} \left( g \mathcal{P}_{k_p} \mathcal{P}_{k_p^\prime} \hat P^\dag_{k_s} \hat P^\dag_{k_i} + {\rm H.c.} \right) \delta_{\bk_s + \bk_i, \bk_p+ \bk_p^\prime},
\end{equation}
where $k=(i,\bk)$ labels the polariton branch and the 2D wavevector. $\mathcal{P}_{k_p}$ and $\mathcal{P}_{k_p'}$ denote the classical amplitudes of the pumped modes and the interaction strength $g$ depends on all the details of the four involved states \cite{g_specification}. Eq.~\eqref{chi3} describes the physical process of two coherent pump polaritons being scattered into a signal-idler polariton pair,  which satisfies the phase matching conditions
\begin{equation}\label{phase-matching}
\left\{ \begin{array}{l}
         \bk_p + \bk'_p = \bk_s + \bk_i \\
         E_{k_p} + E_{k_p^\prime} = E_{k_{s}} + E_{k_{i}}. \end{array} \right.
\end{equation}
The resulting shape of the available states depends on the energy dispersion curves and on the positions of the pump beams. In single planar cavity setups with one~\cite{LangbeinPRB2004} or two pumps~\cite{PortolanEPL09,SavastaPRL2005} the available phase-space reduces to curves where at most two of the final states can have the same energy. 
Multi-mode settings provide a much larger flexibility~\cite{CiutiPRB2004}. For the present planar device and the pump configuration shown in Fig.~\ref{Coupled_cavities} c), the phase-matching conditions~\eqref{phase-matching} are fulfilled on a whole circle of energy-degenerate states in the middle polariton branch. This will allow simultaneously the generation of hyper-entangled states and the reduction of the detrimental phonon-induced noise background as shown below.

{\it Polarization and path entanglement.---}  For pump fields with a definite circular polarization ($\sigma=\pm1$), this polarization will be inherited by the polariton modes and due to spin-preserving Coulomb interactions only one of the four polarization configurations $|+,+\rangle$, $|-,-\rangle$, $|+,-\rangle$ or $|-,+\rangle$ is created. For two linearly co-polarized pump beams all the four polarization states are activated and for the generation of polarization entanglement the only useful configuration occurs for linearly cross-polarized pump fields~\cite{PortolanEPL09}. In this case the counter-circular channel (due to bound biexciton and two-exciton scattering states of opposite spin) is suppressed owing to destructive interference~\cite{suppinfo}. As a result, the generated photon pairs are produced in an entangled state of the form $(|+,+\rangle + |-,-\rangle)/\sqrt{2}$.

By generalizing Eq.~\eqref{chi3} to account  for polarization selection rules and restricted to cross-polarized driving, the photon pair creation process is described by an effective Hamiltonian 
\begin{equation}\label{Heff}
\hat H_{\text{eff} }= \sum_{\bk_s \bk_i} G \left(\hat P^\dag_{\bk_s +} \hat P^\dag_{\bk_i+} + \hat P^\dag_{\bk_s-} \hat P^\dag_{\bk_i-} \right) + {\rm H.c.}, 
\end{equation}
where $G=g \mathcal{P}_{k_{p}} \mathcal{P}_{k_{p}^{\prime}}$ and $\bk_i = \bk_{p} + \bk_{p}^\prime - \bk_s$ is assumed.
For the configuration shown in Fig.~\ref{Coupled_cavities} c) $|\bk_i|=|\bk_s|$ and by selecting specific paths (say $\bk_1$ and $\bk_2$, $-\bk_1$ and $-\bk_2$) the outgoing photon pair is generated in a \textit{hyper-entangled} state
\begin{equation}
|\psi\rangle = \frac{1}{2} {\Big (} |\bk_1,-\bk_1\rangle + |\bk_2,-\bk_2\rangle {\Big )} \otimes {\Big (} |+,+\rangle + |-,-\rangle {\Big )},
\end{equation}
which exhibits entanglement in both the momentum and the polarization degrees of freedom. Being degenerate in energy the photons in modes $\bk_1$ and $\bk_2$ can be interfered, and the momentum can be used as an independent degree of freedom in photonic entanglement experiments~\cite{BarbieriPRA2005}.

{\it Photoluminescence.---} In polariton systems the fidelity of the EPPs is affected by phonon-induced scattering processes and Rayleigh scattering from the pump beams. The effect of Rayleigh scattering is strongly suppressed  as soon as the pump and the signal-idler photons are non energy-degenerate. On the contrary, at temperatures of a few Kelvin, polaritons can scatter incoherently by emission or absorption of acoustic phonons, and redistribute along the dispersion curve. This \textit{pump-induced} photoluminescence (PL) noise background competes with parametric coherent photoemission and lowers the degree of non-classical correlations~\cite{ZeilingerNature2001,SavastaPRL2005}. The multi-cavity setting can reduce its impact by using the mode splitting $J$ to create a large energy separation between the pump and the signal-idler photons.

To quantify the amount of entanglement in the present coupled cavity setup, we study the competition between parametric coherent scattering and the incoherent PL background. Starting from Eq.~\eqref{Heff}
, the dynamics of the signal and idler mode can be evaluated within a two-mode description~\cite{PortolanPRB2008_Langevin}
\begin{eqnarray}\label{H-L}
\partial_t \hat P_{k_s} &=& - i \left( \omega_{k_s}  -i \frac{\Gamma_{k_s}}{2} \right)\hat P_{k_s} -i G \hat P^\dag_{k_i}  + \mathcal{\hat F}_{k_s}, \label{eq:Pdot1} \\
\partial_t \hat P^\dag_{k_i} &=& i \left( \omega_{k_i}  +i \frac{\Gamma_{k_i}}{2} \right) \hat P^\dag_{k_i} +i G^* \hat P_{k_s} + \mathcal{\hat F}^\dag_{k_i}, \label{eq:Pdot2} 
\end{eqnarray}
where the background PL is treated separately from the parametric interaction and enters through the time-dependent Langevin noise operators $\mathcal{\hat F}_{k_s}$ and $\mathcal{\hat F}_{k_i}$. 
Under continuous driving conditions the effect of noise is fully determined by the stationary correlators $\langle  \mathcal{\hat F}^\dag_k(t)  \mathcal{\hat F}_k(t') \rangle = \Gamma_k N^{\text{PL}}_k\delta(t-t') $ and  $\langle  \mathcal{\hat F}_k(t)  \mathcal{\hat F}^\dag_k(t') \rangle = \Gamma_{k} (N^{\text{PL}}_k+1)\delta(t-t') $, where the total polariton decay rates $\Gamma_k$ and the stationary occupations $N^{\text{PL}}_k$ are evaluated in the following.

\begin{figure}
\begin{center}
\includegraphics[width=0.45\textwidth]{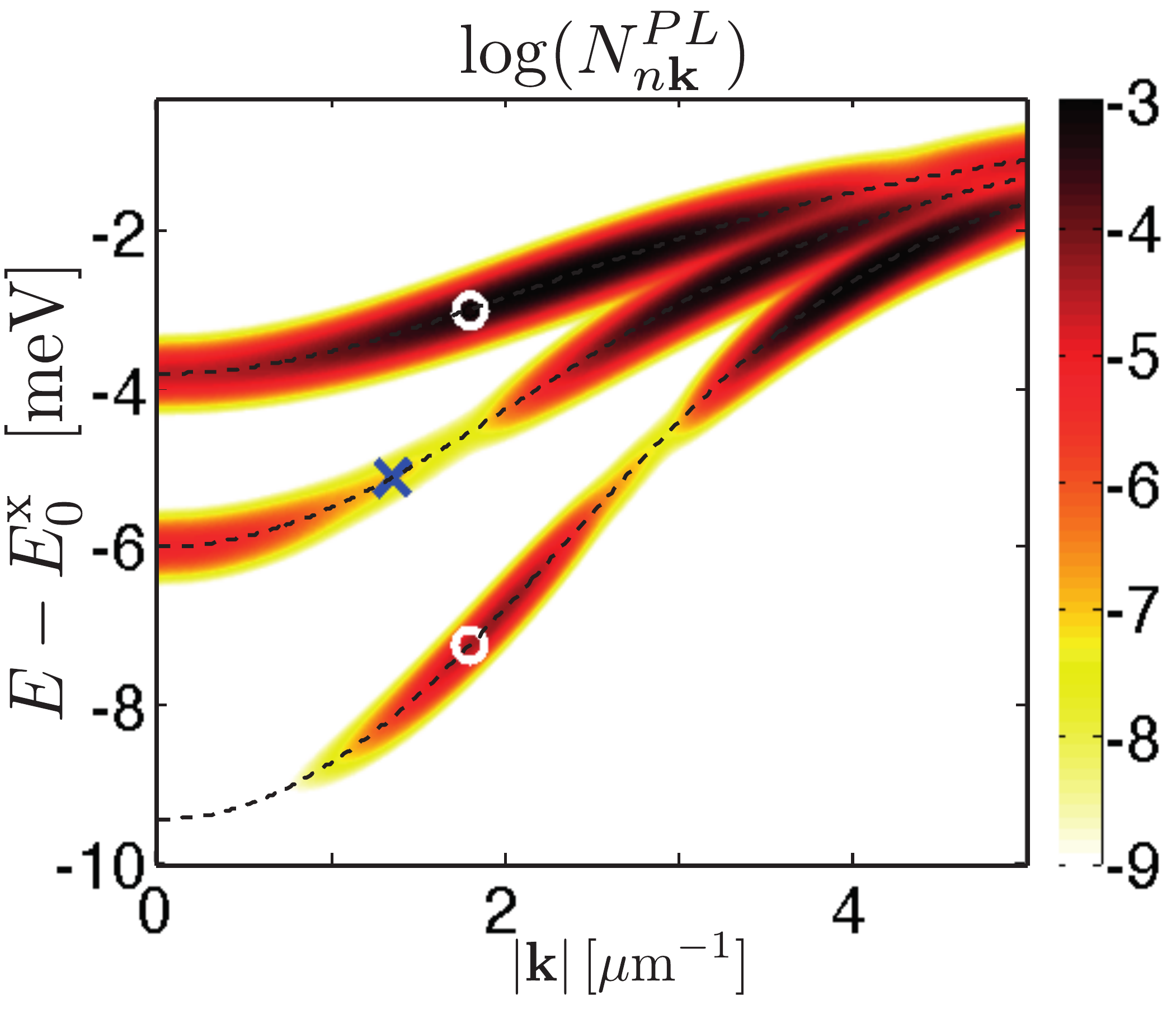}\caption{Stationary polariton populations in a logarithmic scale and superposed on the respective energy dispersion curve. In this line cut of the 2D k-space, the two coherent pump modes are resonant at the white circles, while the blue cross refer to the intersection with the circle of degenerate signal/idler pairs of Fig.~\ref{Coupled_cavities} c). Phonon-induced scattering is quasi-elastic and favors quasi-resonant processes.}\label{SteadyState}
\end{center}
\end{figure}

The dominant  incoherent processes for optically generated polaritons in III-V systems are (acoustic) phonon-induced scattering and radiative losses.  
In strainfree heterostructures, the exciton-phonon interaction is well described by a 3D bulk-like model,  
\begin{equation}\label{DF3well}
\hat H^{\text{DF}}_\text{exc-ph} = \sum_{\bk,\bq,q_z,n} \Xi(\bq,q_z) \hat B^\dag_{n,\bk+\bq}\hat B_{n; \bk} \left(e^{i q_z \lambda_n } \hat b_{\bq,q_z} + {\rm H.c.} \right),
\end{equation}
where $\hat b_{\bq,q_z}$ is the bosonic operator for phonons with transverse momentum $\bq$ and momentum $q_z$ along the confinement axis.
In Eq.~\eqref{DF3well} $\lambda_n = n\ \text{z}_\text{qw}$ accounts for the position of the wells numbered as $n=-1,0,1$ and the coefficient $ \Xi(\bq,q_z)$ contains the exciton overlap integrals in terms of the Fourier transform over the phonon wavevectors \cite{TassonePRB1997,PortolanPRB2008_Langevin}.


Using Fermi's golden rule, the linear PL dynamics for the populations $N^{\text{PL}}_\bk$, for each $\sigma$, can be described microscopically by a Boltzmann equation
\begin{equation}\label{Boltzmann}
\partial_t N^{\text{PL}}_\bk = I_\bk -\Gamma_\bk N^{\text{PL}}_\bk + \sum_{\bk'} W^{(\text{ph})}_{\bk,\bk'} N^{\text{PL}}_{\bk'},
\end{equation}
where the total linewidth $\Gamma_\bk = \Gamma^{(\text{ph})}_\bk + \gamma^{\text{(rad)}}_\bk$ includes phonon-induced and radiative losses.  
Under continuous wave (CW) excitation with two lasers resonant with $E_{\bk_p}$ and $E_{\bk_p'}$, we model the pump term $I_k$ as two Gaussian profiles centered around $\bk_p$ and $\bk_p'$~\cite{suppinfo}. The total phonon-induced polariton scattering rate from state $\bk$ to state $\bk'$ is $W^{(\text{ph})}_{\bk',\bk} = W^{+}_{\bk',\bk} + W^{-}_{\bk',\bk}$,
where the rates from phonon emission $(+)$ and absorption $(-)$ are given by
\begin{equation}\label{eq:Wpm}
\begin{split}
&W^{\pm}_{\bk',\bk} = \frac{1}{\rho
u S} \frac{|\bk'-\bk|^2 +(q^0_z)^2}{|\hbar u q^0_z|} |\Xi (\bk-\bk',q_z)|^2  \\
& \times \left|\sum_p  e^{i q^0_z \lambda(p)} X^p_{k'} X^p_{k}\right|^2 \left(n_B(E^{ph}_{q}) + \frac{1}{2} \pm \frac{1}{2}\right).
\end{split}
\end{equation}
Here $n_B(E)$ is the Bose distribution, $\rho$ is the density and $u$ is the sound velocity of the material and  $S$ is the quantization surface.
In Eq.~\eqref{eq:Wpm} $E^{ph}_{q}$ is the phonon energy and the 3D phonon wave vector is $\vec q=({\bf
q}=\bk-\bk',q^0_z)$, where $q^0_z$ is calculated from the condition of energy
conservation, $\hbar \omega_{{\bk'}} - \hbar\omega_{{\bk}} \pm E^{ph}_{q}=0$. The expression for $\Xi (\bk-\bk',q_z)$ is explicitly derived in~\cite{suppinfo}.
The sum over all final scattering states gives the total phonon-induced loss rate $\Gamma^{(\text{ph})}_\bk = \sum_{\bk'}  W^{(\text{ph})}_{\bk',\bk}$. The radiative linewidth is $\gamma^{\text{(rad)}}_\bk=\sum_n |X^n_\bk|^2 \gamma^n_{\rm c}$, and in the following we take a typical cavity loss rate of $\gamma^n_{\rm c}= 0.35$ meV.

Figure~\ref{SteadyState} shows the stationary polariton occupations $N^{\text{PL}}_\bk$ derived from Eq.~\eqref{Boltzmann} and for other parameters specified above and in~\cite{LangbeinPRB2004}. 
For simplicity, but without loss of generality, a radial symmetry is assumed~\cite{suppinfo} and the population distribution is plotted as a function of $|\bk|$. This is consistent with radial-symmetric steady-state populations observed in experiments \cite{LangbeinPRB2004,PortolanPSSc2008}.   
Phonon-induced polariton scattering favors quasi-elastic events exchanging small energies \cite{TassonePRB1997}. If more branches are present the intra- and inter-branch 
scattering rates will be of the same order and the polariton population will spread across the branches as well. 
The pumped polariton modes in the first and third branch scatter dominantly with low energy exchange into the second branch, which is separated from the other branches by more than 2 meV. Since under the relevant conditions $\gamma^{\text{(rad)}}_\bk \gg \Gamma^{(\text{ph})}_\bk$ multi-phonon processes are highly suppressed, this creates a PL window for the signal and idler wave vectors around $\bk = (1.5,0)\ \mu$m$^{-1}$, where the PL population is reduced by more than 3 orders of magnitude with respect to the intra-branch PL on the third branch.

\begin{figure}
\begin{center}
\includegraphics[width=0.45\textwidth]{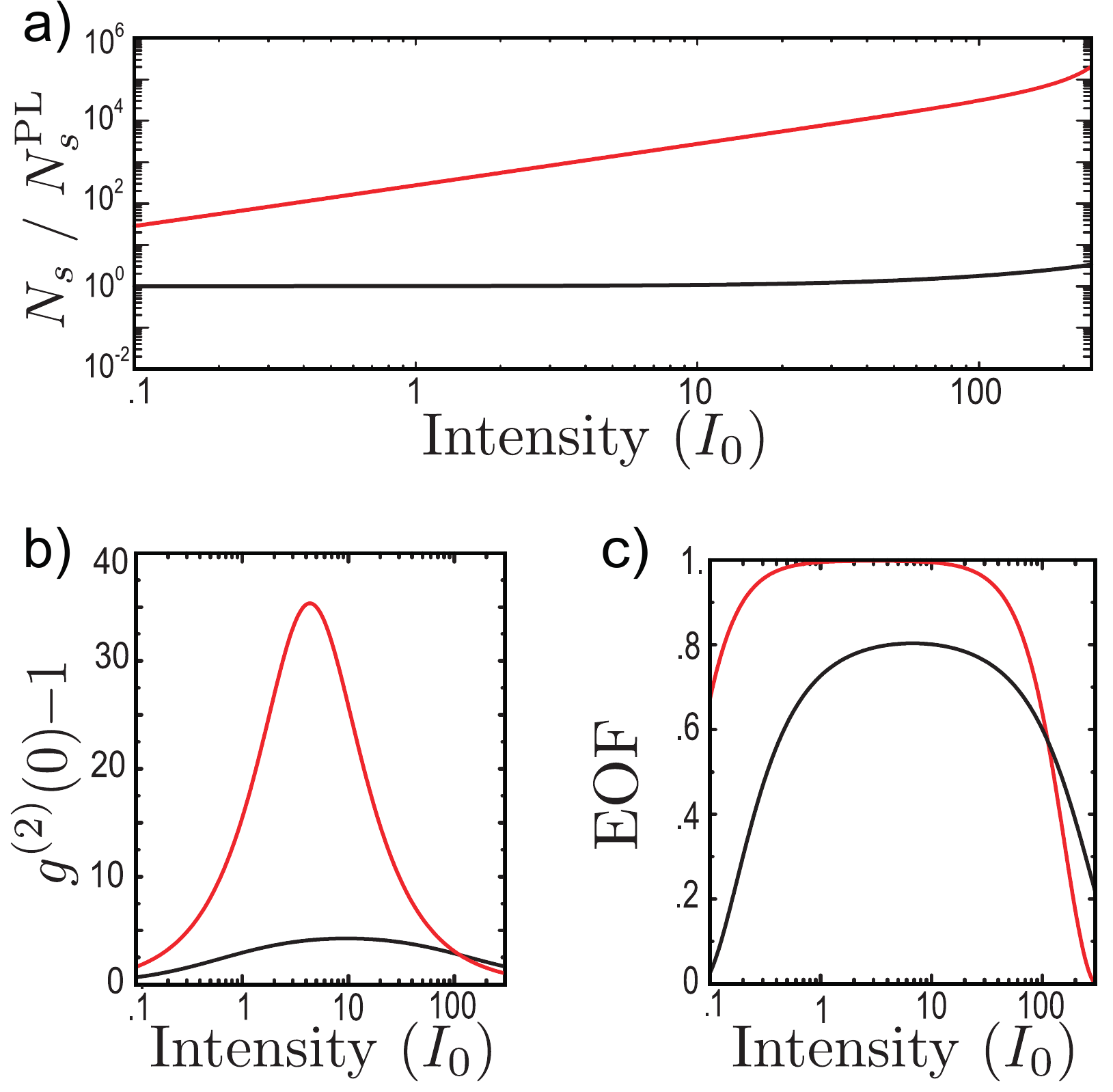}\caption{(color online) Single (black) - triple (red) cavity comparison in steady-state. a) Signal to noise ratio, defined as total generated signal $N_s$ over background PL $N^{PL}_s$ at $\bk = 1.5\ \mu$m$^{-1}$. b) Equal-spin autocorrelation at zero delay $g^{(2)}(0)$. c) Theoretically achievable EOF. In the simulation, a pump independent background noise of $N_s^{0}=10^{-5}$ has been added to account for other noise sources in the system and the reference pump intensity $I_0$ corresponds to an injected polariton density of about 0.5 $\mu$m$^{-2}$ \cite{LangbeinPRB2004}. }\label{3vs1}
\end{center}
\end{figure}
%
%


{\it Entanglement quantification.---} To provide a direct comparison between the present multi-mode setup and entanglement generation schemes using only a single microcavity, we evaluate in Fig.~\ref{3vs1} three different quantities that are of importance for the experimental verification of entanglement in these systems.  In Fig.~\ref{3vs1} a) we first plot the signal to noise ratio (SN) defined as the total generated signal $N_s$ over the background PL $N^{PL}_s$ at the detected signal and idler wavevectors. The red line shows the results obtained from the steady state solution of Eqs.~\eqref{eq:Pdot1} and~\eqref{eq:Pdot2} and the PL noise level evaluated with Eq.~\eqref{Boltzmann}. The black line indicates the corresponding results for a single-cavity setup \cite{PortolanEPL09}, but with otherwise identical parameters.  
We  see that for a single cavity, the emitted photons are dominated by the PL background, whereas in the three-cavity setting the coherent signal is clearly above the PL noise level and dominated by the parametric (pair) emission. 

Under the same conditions we evaluate in Fig.~\ref{3vs1} b) and c) the equal time two-photon correlation function $g^{(2)}(0)=\langle \hat P^\dag_s\hat P^\dag_{i} \hat P_{i} \hat P_{s} \rangle/(\langle\hat P^\dag_s \hat P_s\rangle\langle\hat P^\dag_i \hat P_i\rangle) $ and 
the entanglement of formation (EOF) \cite{Wootters_1997_and_1998} as two measures 
of signal-idler correlations and of the degree of polarization entanglement~\cite{Edamatsu,Senellart_Nature2010} between two simultaneously emitted photons, respectively. The EOF is evaluated from the reconstructed two-polariton subsector of the full density operator~\cite{Tomography,Portolan_bookchapter2009}, which can be obtained experimentally from measurements of all the four-operator expectation values $\langle \hat P^\dag_{s \sigma}(t_1) \hat P^\dag_{i \sigma'}(t_2) \hat P_{i \bar \sigma}(t_2) \hat P_{s \bar \sigma'}(t_1) \rangle$.   
In contrast to the total signal, the quantities plotted in Fig.~\ref{3vs1} b) and c) are sensitive to detected photon pairs only and consistent with previous findings~\cite{PortolanEPL09} we observe bunching, $g^{(2)}(0)>1$, and polarization entanglement ${\rm EOF} >0$ for both configurations. However, 
the bunching of $g^{(2)}(0)\sim4$ achievable with the single-cavity set-up is at the border of the acquisition capabilities of current experiments \cite{Edamatsu}, while the triple cavity structure shows a ten-fold enhanced effect, $g^{(2)}(0) > 30$, in relation with a very high SN ratio.
Moreover, there is a pronounced difference in the fidelity of the entangled photons pairs, and the ${\rm EOF}$ in the three-cavity setting can be close to the value of ${\rm EOF}=1$ expected for a pure polarization entangled Bell state.


{\it Conclusions.---} In this work we showed that coupled microcavity structures can become an innovative design for protecting 
quantum coherences from solid-state background noise and we applied this concept to devise innovative bright sources of entangled and hyper-entangled photon pairs. 
Our analysis predicts a suppression of the phonon-induced noise of more than 3 orders of magnitude, which thereby eliminates one of the main noise sources in current experiments.

\begin{acknowledgments}
S.P. thanks S. Savasta and F. Rossi for many stimulating discussions. This work was supported by the EU project SIQS and the Austrian Science Fund (FWF) through SFB FOQUS and the START grant Y 591-N16.
\end{acknowledgments}




\cleardoublepage
\begin{widetext}

\begin{center}
{\large \textbf{Supplementary Material for:\\
Generation of hyper-entangled photon pairs in coupled microcavities}}

\bigskip
\bigskip


{S. Portolan$^1$, L. Einkemmer$^2$, Z. V\"or\"os$^2$, G. Weihs$^2$, P. Rabl$^1$ \\}
\textit{$^1$ Institute of Atomic and Subatomic Physics, TU Wien, Stadionalle 2, 1020 Wien, Austria \\}
\textit{$^2$ Institut f\"ur Experimentalphysik, Universit\"at Innsbruck, Technikerstr. 25, 6020 Innsbruck, Austria}
\end{center}

\bigskip
\bigskip

In this supplemental material we provide a more detailed derivation of the different results presented in the main part of the paper. 
To be consistent with our notation, we start in Sec.~\ref{sec:Model} with a brief review of the standard exciton-polariton model. In Sec.~\ref{sec:ThirdOrder} we detail the derivation of the effective $\chi^{(3)}$ nonlinear Hamiltonian given in Eq. (3) in the main text, taking into account both the Coulomb interaction between excitons as well as the effective nonlinearity arising from the non-bosonic nature of the excitons. In Sec.~\ref{sec:Polarization} we discuss the generation of polarization entangled photons by using cross-polarized pumping beams and in Sec.~\ref{sec:Phonons} we present the explicit form of the acoustic phonon interaction Hamiltonian for the three-cavity setup. Finally, Sec.~\ref{sec:Numerics} contains additional details on the numerical simulation of the photoluminescence signal presented in Fig. 2 of the main text.

\section{Model} \label{sec:Model} 
Polaritons are mixed quasiparticles resulting from the strong coupling between light and electronic excitations
(excitons) in semiconductor crystals. For the quasi-2D interacting electron system we adopt the usual
semiconductor model Hamiltonian \cite{SM_AxtKuhn}. We shall consider III-V direct-gap materials, for which a  two-band semiconductor model well reproduces the optical response near the band edge. It reads:
\begin{equation}\label{2bands} 
\hat{H}_e = \hat{H}_0^{eh} + \hat{H}_{\text{C}}\, . 
\end{equation}
The first term describes the single-particle Hamiltonian terms for electrons in
conduction band and holes in valence band \begin{equation}\label{H
zero} \hat{H}_0^{eh}  = \sum_{\bf k} E_{c,{\bf k}} \hat{c}^\dag_{{\bf
k}} \hat{c}_{{\bf k}} + \sum_{\bf k} E_{h,{\bf
k}}\hat{d}^\dag_{{\bf k}} \hat{d}_{{\bf k}},
\end{equation}
where $\hat{c}^\dag_{{\bf k}}$ ($\hat{d}^\dag_{{\bf k}}$) creates an electron (hole) in the conduction (valence) band with a 2D quasi momentum ${\bf k}$. The second term in Eq.~\eqref{2bands} is the Coulomb interaction,  
\begin{eqnarray}\label{ V coul} \hat{H}_{\text{C}} && = \frac{1}{2} \sum_{{\bf q}
\neq 0} \sum_{{\bf k},{\bf k}'} V_q \hat{c}^\dag_{{\bf k}+{\bf q}}
\hat{c}^\dag_{{\bf k}'-{\bf q}} \hat{c}_{{\bf k}'}
\hat{c}_{{\bf k}} + \frac{1}{2} \sum_{{\bf q} \neq 0} \sum_{{\bf
k},{\bf k}'} V_q \hat{d}^\dag_{{\bf k}+{\bf q}}
\hat{d}^\dag_{{\bf k}'-{\bf q}} \hat{d}_{{\bf k}'}
\hat{d}_{{\bf k}} - \nonumber \\
&& - \sum_{{\bf q} \neq 0} \sum_{{\bf k},{\bf k}'} V_q
\hat{c}^\dag_{{\bf k}+{\bf q}} \hat{d}^\dag_{{\bf k}'-{\bf q}}
\hat{d}_{{\bf k}'}c_{{\bf k}}\,, 
\end{eqnarray}
where the three contributions represent the repulsive electron-electron (e-e) and
hole-hole (h-h) terms and the attractive (e-h) interaction, respectively.

Hamiltonian (\ref{2bands}) can be rewritten
as
\begin{equation}\label{Ham electron} \hat{H}_e = \hat{H}_0^{eh}  +
\hat{H}_{\text{C}}=
\sum_{N \alpha} E_{N \alpha} \ket{E_{N \alpha}} \bra{E_{N \alpha}}\,,
\end{equation} 
where the eigenstates of $\hat{H}_e$, with energies
$E_{N \alpha} =\hbar \omega_{N \alpha}$, have been labeled
according to the number $N$ of {\em eh} pairs. The state
$\ket{E_{N=0}}$ is the electronic ground state, the $N=1$ subspace
is the one exciton subspace with the collective quantum
number $\alpha$ denoting the exciton energy level $\nu$, the in-plane
wave vector ${\bf k}$ and the spin index $\sigma$.  Throughout this work we are
interested in studying polaritonic effects, where the optical
response involves mainly excitons belonging to the 1S band with wave
vectors close to normal incidence, $|{\bf k}| \ll \frac{\pi}{a_{\text{x}}}$, ${a_{\text{x}}}$ being the exciton Bohr radius, typically around $10$ nm.
In the following we will omit the internal quantum number $\nu$. 



Eigenstates of the model Hamiltonian with $N=1$ (called excitons) can
be created from the ground state by applying the exciton creation
operator,
\begin{equation}
\bigl| N=1, \sigma,{\bf k}\bigr>=\hat{B}^\dagger_{\sigma,
{\bf k}} \bigl| N=0\bigr> \ \ , \ \ \hat{B}^\dagger_{
\sigma, {\bf k}} = \sum_{{\bf k}'} \Phi^{\bf k}_{\sigma, {\bf
k}'} \hat{c}^\dagger_{\sigma, {\bf k}' + \eta_e{\bf
k}}\hat{d}^\dagger_{\sigma, -{\bf k}' + \eta_h{\bf k}}\, ,
\label{Bdag}
\end{equation}
where $\Phi^{\bf k}_{\sigma {\bf k}'}$ is the 1S exciton wave function of total and relative wave
vectors ${\bf k}$ and ${\bf k}'$, respectively.  The coordinate transformation from elecron-hole to center-of-mass and relative momenta read ${\bf k} = {\bf k}_e - {\bf k}_h$,  ${\bf k}' = \eta_h{\bf k}_e +
\eta_e{\bf k}_h$ ~\cite{SM_Sham_PRB1998,PortolanPRB2008_DCTS}, where $\eta_{(e,h)}=m_{(e,h)}/(m_{(e,h)}+m_{(h,e)})$ and $({\bf k}_e,{\bf k}_h)$ are the electron and hole wave
vectors.) For the multicavity setup with $n_{\rm cav}$ coupled microcavities, we label excitons in different quantum well by an additional index $n=1,\dots, n_{\rm cav}$, i.e. $ \hat{B}^\dagger_{\sigma,
{\bf k}}\rightarrow \hat{B}^\dagger_{n,\sigma,
{\bf k}}$ and whenever needed we adopt the following shorthand notation $k\equiv (n,\sigma,{\bf k})$.

Restricted to a single transverse cavity mode, the free Hamiltonian for photons in the coupled cavity array is given by
\begin{equation}\label{Ham cavity} 
\hat{H}_c =
\sum_{n=1}^{n_{\rm cav}}\sum_{{\bf k}} \hbar \omega_{n{\bf k}} \hat{a}_{n\bf k} ^{\dag} \hat{a}_{n\bf k}- \sum_{n=1}^{n_{\rm cav}-1} \sum_{\bk} J_{n,n+1}\left( \hat a^\dag_{n \bk} \hat a_{n+1 \bk} + {\rm H.c.} \right)\,.
\end{equation} 
Here $\omega_{n{\bf k}}= \sqrt{(\omega_{n}^0)^2+ (c/n_{\rm ref})^2 |\bk|^2}\approx \omega_{n}^0+ (c/n_{\rm ref})^2 |\bk|^2/2$  \cite{SM_Savona Quattropani SSC}  is the quadratic dispersion of the individual cavities with refractive index $n_{\rm ref}$ and $J_{n,n+1}$ is the mode coupling between neighboring cavities, which depends on the transmission of the Bragg mirrors. The discrete modes of the outermost cavities are coupled to the external continuum of modes, which can be described within a quasi-mode theory~\cite{SM_PortolanPRB2008_DCTS} by an effective Hamiltonian
\begin{equation}\label{H quasi modi} \hat{H}_p = i \hbar \,  \sum_{j=1,n_{\rm cav}} \sum_{\bf k} t_{j} (\hat{E}^{(-)}_{j,\bf k} \hat{a}^\dag_{\bf k} -
\hat{E}^{(+)}_{j,\bf k} \hat{a}_{\bf k})\, ,\end{equation} where $t_j$ determines the fraction of the transmitted
field amplitude at mirror $j$, and $\hat{E}^{(-)}_{j,\bf k}$
($\hat{E}^{(+)}_{j,\bf k}$) is the positive (negative) frequency
part of the external field.

The dipole coupling of the electron
system to the cavity modes is given within the usual rotating wave
approximation by
\begin{equation}\label{Ham inter cav-exc} 
\hat{H}_{I} = \sum_{n {\bf k}}
\hbar V_{n {\bf k}} \hat{a}_{n\bf k} ^{\dag} \hat{B}_{n {\bf k}} +
{\rm H.c.}\, , \end{equation}
where $V_{nk}$ is the
photon-exciton coupling coefficient enhanced by the presence of the
cavity \cite{SM_Savona Quattropani SSC}.  $V_{n\bk}$ is proportional to the overlap between the exciton wave function $\Phi^{\bf k}_{n\sigma {\bf k}'}$ and the transverse optical modefunction in cavity $n$, which can be optimized by an appropriate design of the structure. 

%

\section{Polariton $\chi^{(3)}$-nonlinearity}\label{sec:ThirdOrder}

Many descriptions of polariton parametric processes make use of the picture of polaritons as interacting bosons \cite{SM_Ciuti parlum,SM_Ciuti parAmpl, SM_LosannaCC,SM_LangbeinPRB2004}.
In these models bosonic commutation relations are assigned to the exciton operators $\hat{B}_{n {\bf k}}$ and $\hat{B}_{n {\bf k}}^\dag$ and an effective boson-boson coupling is added to the Hamiltonian to account for the non-bosonic exciton character (``phase-space filling''). Combined with the Coulomb interaction this coupling is then treated within the mean-field approximation. At low polariton densities this approach provides a good description of the resulting nonlinear third order response and is often used due to its simplicity and its direct analogy to nonlinear optics. 
However, in particular at higher densities the absolute strength of the nonlinear terms differs from those obtained from more rigorous microscopic theories \cite{SM_PortolanPRB2008_DCTS,SM_EinkemmerPRB2013}. For all the result presented in the main part of this paper these microscopic corrections -- as described in the following -- have been taken into account. 

\subsection{Heisenberg equations of motion}

In the case of a single cavity and a single pump field, the Heisenberg equations of motion for a polariton system up to the third-order nonlinearity have been fully worked out in Ref.~\cite{SM_PortolanPRB2008_DCTS,SM_PortolanPRB2008_Langevin}.  Taking these references as our starting point, we adapt here this analysis for the case of multiple coupled cavities. For concreteness we address directly the triple cavity case ($n_{\text{cav}}=3$), but the following line of argument is completely general. We assume that the first cavity is driven by two coherent pump fields, which resonantly excite the polariton modes with wave vectors ${\bf k}_p$ and ${\bf k}_{p'}$, respectively. 
%
%
%
%
%
Following the same notation as in the paper, we introduce a vector ${\mathbf{B}}_{\bk}$ with components 
$
{\mathbf{B}}_{\bk} = ( \hat{B}_{1\bf k}, \hat{a}_{1 \bf k},\hat{B}_{2\bf k}, \hat{a}_{2 \bf k}, \hat{B}_{3\bf k}, \hat{a}_{3 \bf k})^T
$Ê and write the resulting equations of motion in a compact form as
\begin{equation}\label{dif1}
\dot {\mathbf{B}}_{\bk} =
-i\mathbf{ \Omega}^{\text{xc}}_{\bk}\,
\mathbf{B}_{\bf k} + \mathbf{E}^{in}_{\bf k}
-i \mathbf{{R}}_{\bf k}^{NL}.
\end{equation}
Here  the matrix 
\begin{equation}
\mathbf{\Omega}^{\text{xc}}_{\bf k} \equiv \left( \begin{array}{cccccc}
\omega^\text{x}_{1 \bf k} &  V_1 & & & & \\
V_1 &\omega^c_{1 \bf k} &  & -J_{1,2} & & \\
 & &\omega^\text{x}_{2 \bk} & V_2 & & \\
 & -J_{1,2} & V_2 & \omega^c_{2 \bk} & & -J_{2,3} \\
 & & & &\omega^\text{x}_{2 \bk} & V_3 \\
  & & & -J_{2,3} &V_3 & \omega^c_{2 \bk} 
\end{array} \right),
\end{equation}
described the linear dynamics of the coupled modes and the vector $\mathbf{E}^{in}_{\bf k}$ accounts for the external driving field. For the pump configuration specified above  $\left( \mathbf{E}^{in}_{\bf k}  \right)_2= t_{1} \left({E}^{in}_{p} e^{-i\omega_p t}\delta_{\bf k,\bf k_p}+{E}^{in}_{p'} e^{-i\omega_{p'} t}\delta_{\bf k,\bf k_{p'}}\right) $ and $\left( \mathbf{E}^{in}_{\bf k}  \right)_i=0$ otherwise. The last term in Eq.~\eqref{dif1}, where 
$
\mathbf{{R}}_{\bf k}^{NL} = ( \hat{R}^{NL}_{1\bf k}, 0,\hat{R}^{NL}_{2\bf k}, 0, \hat{R}^{NL}_{3\bf k}, 0)^T
$,Ê 
describes additional nonlinear contributions, which we address in the following. Note that in Eq.~\eqref{dif1} we have for simplicity omitted the photon loss terms, which can be included by adding photon loss rates for the cavity operators $\hat a_{n{\bf k}}$ and the corresponding Langevin noise operators.


%

Coulomb and photon-exciton interactions will be effective only between exciton and photon mode belonging to the same cavity. As a consequence the relevant nonlinear source terms, which couple waves with different in-plane wave vector ${\bf k}$, can be treated very similarly to the single cavity case. We write the nonlinear terms as $\hat{ R}^{NL}_{n \bf k}=(\hat{ R}^{\rm sat}_{n \bf
k}+\hat{R}^{\text{xx}}_{n \bf k})$. The first term
originates from the phase-space filling of the exciton transition and couples excitons with the same spin,
\begin{equation}\label{Rsat}
\hat{R}^{\rm sat}_{n {\bf k}} = \frac{V}{n_{\text{sat}}} \sum_{\bk' \bk'' }\hat B_{n{\bf k'}} \hat a_{n
{\bf k''}} \hat{B}^\dag_{n \bar{\bf k}}\, ,
\end{equation}
where ${n_{\rm sat}}= 7/(16 \pi a^2_\text{x})$ is the exciton saturation density \cite{SM_Savasta_SST2003,SM_PortolanPRB2008_DCTS} and $\bar{\bf k}= {\bf k^\prime}+{\bf k''}-{\bf k}$.  The second term comes from the Coulomb interaction among electrons and holes and contains two contributions. In the co-circular channel, only particles with the same spin are involved and the resulting interaction is always repulsive. In contrary, the counter-circular term describes the scattering of excitons of opposite spin and it can include a bound biexciton intermediate state. As a consequence its strength and its sign can vary considerably around the biexciton binding energy~\cite{SM_Schumacher_PRB2007}. Altogether, and using the same notation as in Ref.~\cite{SM_Schumacher_PRB2007},  we obtain
\begin{eqnarray}\label{XX}
\hat{R}^{\text{xx}}_{n {\bf k}}(t)  & = &\sum_{\bk',\bk''} \hat{B}^\dag_{n\sigma \bar{\bf k}}(t) \int_{-\infty}^t dt' \
\tilde T^{++}(t-t') \hat B_{n \sigma {\bf k'}}(t') \hat B_{n \sigma {\bf k''}}(t') \nonumber \\
&& - \sum_{\bk',\bk''} \sum_{\sigma'=\pm} \hat{B}^\dag_{n -\sigma' \bar {\bf k}}(t) \int_{-\infty}^t dt' \
\tilde T^{+-}(t-t') \hat B_{n \sigma' {\bf k'}}(t') \hat B_{n -\sigma' {\bf k''}}(t')\, .
\end{eqnarray}
The transition T-matrix includes the instantaneous mean-field exciton-exciton
interaction contribution and a non-instantaneous term originating from
four-particle correlations \cite{SM_Savasta_SST2003,SM_Schumacher_PRB2007,SM_PortolanPRB2008_DCTS}. The memory-less equal-time limit of Eq.~\eqref{XX} corresponds to an energy independent $\chi^{(3)}$ nonlinearity for excitons, which is usually assumed in effective models, where excitons are treated as interacting bosons. As we show below, within our microscopic theory and in a quasi stationary regime the resulting $\chi^{(3)}$-interaction between polaritons can be derived more rigorously by using a Weisskopf-Wigner approximation for the integral kernel.
%
The transition T-matrix includes the instantaneous mean-field exciton-exciton
interaction contribution and a non-instantaneous term originating from
four-particle correlations. The same notation as in Ref.~\cite{SM_Schumacher_PRB2007} has been used. 

\subsection{Strong coupling and polariton-polariton interactions}

When the coupling rate $V$ exceeds the  decay rate of the exciton
coherence and of the cavity field, the system enters the strong
coupling regime. In this regime, cavity-polaritons arise as the two-dimensional eigenstates of
$\mathbf{\Omega}^{\text{xc}}_{\bf k}$. 

In order to obtain  the dynamics for the polariton system we diagonalise the linear subproblem of Eq. (\ref{dif1}) by a (unitary) Hopfield transformation~\cite{SM_BassaniNuovoCimento1986} $ \mathbf{P}_{\bf k} = \mathbf{U}_{\bf k}\mathbf{B}_{\bf k}$, where $\left(\mathbf{\cal P}_{\bf k}\right)_i=\hat P_{i\bk}$, and express excitons and photons in terms of a set of polaritonic quasi-particle operators 
 \begin{equation}
 \hat P_{i \bk} = \sum_n X^n_{i \bk} \hat B_{n \bk}+ C^n_{i \bk} \hat a_{n \bk},
 \end{equation}
where $X^n_{i \bk}$ and $C^n_{i \bk}$ can be seen as the exciton and photon components of the $n$-th cavity on the $i$-th polariton branch.
By applying this transformation to Eq. (\ref{dif1}) we obtain
%
\begin{equation}\label{dif2}
 {\partial_t \mathbf{P}}_{\bk,\sigma}=-i
 \omega_{\bk}
\mathbf{P}_{\bk,\sigma} + \mathbf{\tilde {E}}^{in}_{\bk,\sigma}
-i \mathbf{\tilde {R}}_{\bk,\sigma}^{NL},
\end{equation}
where the $\omega_{\bk}$ are the polariton eigenfrequencies, $\mathbf{\tilde {R}}^{NL}_{\sigma \bk}=\mathbf{U}_{\bf k} \mathbf{ {R}}^{NL}_{\sigma \bf k}$ and $\mathbf{\tilde {E}}^{in}_{\sigma \bf k}=\mathbf{U}_{\bf k} \mathbf{E}^{in}_{\sigma \bf k}$ .

Let us consider a situation where the energies of the exciting pulses are all close to the corresponding polariton resonance values $\omega_{i \bk}$ and the broadenings are small compared to the splitting between the polariton branches. 
By adopting a Weisskopf-Wigner approximation~\cite{SM_Savasta_SST2003} we can simplify the memory integral in Eq.~\eqref{XX} in the case of continuous wave (CW) excitation and express the result in terms of polariton operators $\hat P_{i\bk}$.
The nonlinear interaction terms we obtain are given by $   (\mathbf{\tilde {R}}^{NL}_{\sigma \bf k})_{i}=\sum^3_{n=1}  \left ((\mathbf{\tilde {R}}^{\rm sat}_{n \sigma \bf k})_{i} + (\mathbf{\tilde {R}}^{\text{xx}}_{n \sigma \bf k})_{i} \right)$ with
\begin{eqnarray}\label{nonlinear1}
(\mathbf{\tilde {R}}^{\rm sat}_{n \sigma \bf k})_{i} &=& \sum_{\bk',\bk''} \sum_{l,m,r} \hat{P}^\dag_{m \sigma \bar{\bf k}}(t)  \left[ X^n_{i \bk} X^n_{m \bar{\bk}} \frac{V}{n_{\text{sat}}} X^n_{l \bk'}C^n_{r \bk''} \right]  \\
  (\mathbf{\tilde {R}}^{\text{xx}}_{n \sigma \bf k})_{i} &=& \sum_{\bk',\bk''} \sum_{l,m,r} \hat{P}^\dag_{m \sigma \bar{\bf k}}(t) \left[ X^n_{i \bk} X^n_{m \bar{\bk}} T^{++}(\omega_{\bk'}+\omega_{\bk''}) X^n_{l \bk'}X^n_{r \bk''} \right]  \hat P_{l \sigma {\bf k'}}(t) \hat P_{r \sigma {\bf k''}}(t) + \label{nonlinear2} \\ 
  &+& \sum_{\bk',\bk''} \sum_{\sigma'=\pm} \sum_{l, m,r} \hat{P}^\dag_{m, -\sigma' \bar{\bf k}}(t) \left[ X^n_{i \bk} X^n_{m \bar{\bk}} T^{+-}(\omega_{\bk'}+\omega_{\bk''}) X^n_{l \bk'}X^n_{r \bk''} \right]  \hat P_{l \sigma' {\bf k'}}(t) \hat P_{r, -\sigma' {\bf k''}}(t), \nonumber
\end{eqnarray}
where $T^{s}(\Omega) = \int^\infty_{-\infty} \tilde T^{s}(\tau) e^{-i\Omega \tau} d\tau$ is the Fourier transform of the time-dependent kernel in Eq.~\eqref{XX} for $s=++,+-$. This approach, which is valid under quasi-stationary conditions, fully accounts for the energy dependence of the exciton-exciton scattering by including in the above equations the frequency dependence of the $T$-matrix.  It can be shown \cite{SM_Savasta_SST2003} that the co-circular channel can be written as the sum of a constant mean-field term $V_{xx}$ plus a genuine four-particle contribution, $T^{++}(\omega) = V_{\text{xx}} + F(\omega)$. In the present work we are interested CW driving and in the scattering of polaritons with a pre-specified energy. In this case the frequency dependence of $T^{++}(\omega)$ simply leads to a renormalization of the coherent scattering amplitude.  In order to maintain a connection with the literature, in the following we will use loosely $V_{\text{xx}}$ as the Coulomb-induced co-circular interaction including its correction beyond mean-field. 

In the last step we now assume that the two pumped polariton modes are strongly driven and we replace the corresponding operators by their classical expectations values, $\hat P_{k_p} \rightarrow \mathcal{P}_{k_p}= \langle \hat P_{k_p}\rangle$ and $\hat P_{k_{p'}} \rightarrow \mathcal{P}_{k_{p'}}= \langle \hat P_{k_{p'}}\rangle$, where again the convention  $k_p=(i_p,\bk_p,\sigma_p)$ is assumed. As discussed below, we are mainly interested in cross-polarized pump beams to get rid of the spin non-conserving scattering channel by destructive interference. However, for the moment we will leave the polarization of the pump fields unspecified.
%
%
Then, by retaining only the most relevant contribution $\sim  \mathcal{P}_{k_p} \mathcal{P}_{k_{p'}}$ in the expressions given in Eqs.~\eqref{nonlinear1} and ~\eqref{nonlinear2}
%
%
the equation of motion for the generic polariton operator reads
\begin{equation}\label{dtP_2pump}
\frac{d}{dt} \hat{P}_{i \sigma  {\bf k}} = - i \omega_{\bf k}
\hat{P}_{i \sigma  {\bf k}} - i \!\! \sum_{\sigma_{p}, \sigma_{p'}, \bar\sigma}\hat{P}^\dag_{m \bar \sigma \bar{\bf k}} \Big( g_{\sigma {\bf k}}\! \ \mathcal{P}_{{i}_{p}\sigma_{p}  {\bf k}_{p}}\mathcal{P}_{ {i}_{p'} \sigma_{p'} {\bf k}_{p'}} \Big) \, ,
\end{equation}
where $g_{\sigma \bk}$ is a short notation for the coupling strength, which depends on all the $\sigma$'s and $\bk$'s in Eq.~\eqref{dtP_2pump} and is detailed below. This equation of motion can also be obtained in the Heisenberg picture by the effective Hamiltonian $H_{\chi^{(3)}}$ (Eq. (3) in the paper). 
The branch index $m$ and wave vector $\bar \bk$ are defined by the wave vector and energy conservation rules,
\begin{equation}\label{SM_phase-matching}
\left\{ \begin{array}{l}
         \bk_{p} + \bk_{p'} = \bk + \bar \bk \\
         E_{i_{p},\bk_{p'}} + E_{i_{p'},\bk_{p'}} = E_{i,\bk} + E_{m,\bar \bk}. \end{array} \right.
\end{equation}
The nonlinear coupling strength can be written as the sum of the co-circular and counter-circular terms, $g_{\sigma {\bf k}} = g^{++}_{\sigma {\bf k}} + g^{+-}_{\sigma {\bf k}} $, and it includes all the spin selection rules. They read
\begin{eqnarray}\label{g_2pump}
g^{++}_{\sigma {\bf k}} &=& \sum^{n_{\text{cav}}}_{n=1} \Bigg( \delta_{\bar \sigma,\sigma} \delta_{\sigma_p,\sigma_{p'}} \delta_{\sigma_p,\sigma} X^n_{i \bk} X^n_{m \bar \bk} \left[ \frac{V}{n_{\text{sat}}} \left( X^n_{i_{p} \bk_{p}} C^n_{i_{p'} \bk_{p'}} + X^n_{i_{p'} \bk_{p'}} C^n_{i_{p} \bk_{p}} \right) + 2\, T^{++} X^n_{i_{p} \bk_{p}} X^n_{i_{p'} \bk_{p'}}  \right] \Bigg) \nonumber \\
g^{+-}_{\sigma {\bf k}} &=& \sum^{n_{\text{cav}}}_{n=1} \Bigg( \delta_{\bar \sigma,-\sigma} \delta_{\sigma_p,-\sigma_{p'}} X^n_{i \bk} X^n_{m \bar \bk} 2\, T^{+-} X^n_{i_{p} \bk_{p}} X^n_{i_{p'} \bk_{p'}} \Bigg).
\end{eqnarray}


\section{Perpendicular polarization: spin-conserving channels}\label{sec:Polarization}

As the calculation of Ref.~\cite{SM_Schumacher_PRB2007} shows, the +- kernel
in the four-particle nonlinear response does not become 
negligible even in a very negatively detuned region, but  it retains a value of approximately one third of the ++ channel.
However, with two linearly cross-polarized pumps, we can show that the counter-circular channel (due to bound biexciton and two-exciton scattering states of opposite spin) is suppressed owing to destructive interference.

We choose two pump fields with polarization vectors $\hat{e}_{p} = \theta$ and
$\hat{e}_{p'} = \theta + \pi/2$ (e.g. $\theta=0$ means the first
pump polarized along $\hat{x}$, while the second one along
$\hat{y}$). From the equation for the pumped polariton mode
\begin{equation}
\frac{d}{dt} \mathcal{P}_{\sigma_p l_p{\bf k}_p} = - i (\omega_{{\bf
k}_p} - i \Gamma_p/2) \mathcal{P}_{i_p \sigma_p {\bf k}_p} + t_c C^n_{i_p \bk_p}
E^{(+)}_{{\bf k}_p} ( \hat{e}_p \cdot \hat{\sigma})\nonumber\, ,
\end{equation}
and $\hat e_p = \cos \theta
|x\rangle + \sin \theta |y\rangle=2^{-1/2}(e^{i\, \theta} |-\rangle
+ e^{-i\, \theta} |+\rangle)$, 
we can see that any linearly polarized pump excites two circularly
polarized polariton modes, 
\begin{equation}\label{relations}
\begin{array}{cc}
  \mathcal{P}^{(+)}_1(t) = \mathcal{P}_{k_p}(t) \frac{e^{-i \theta}}{\sqrt{2}},
  & \hspace{1.0cm}\mathcal{P}^{(+)}_2(t) = \mathcal{P}_{k_{p'}}(t) \frac{e^{-i \theta}}{\sqrt{2}} e^{-i \pi/2}, 
  \\ [3.5ex]
    \mathcal{P}^{(-)}_1(t) = \mathcal{P}_{k_p}(t) \frac{e^{+i \theta}}{\sqrt{2}},
  & \hspace{1.0cm}\mathcal{P}^{(-)}_2(t) = \mathcal{P}_{k_{p'}}(t) \frac{e^{+i \theta}}{\sqrt{2}} e^{+i \pi/2}. 
\end{array}
\end{equation}
By working out explicitly the spin sum in Eq. (\ref{dtP_2pump}) we obtain

\begin{eqnarray}\label{dtP_2pump_2}
\frac{d}{dt} \hat{P}_{i \sigma  {\bf k}} &=& - i \omega_{\bf k}
\hat{P}_{i \sigma  {\bf k}} - i  g_{\sigma \bk} \hat{P}^\dag_{m \bar \sigma  \bar{\bf k}}   \big(\mathcal{P}^{(+)}_1 + \mathcal{P}^{(-)}_1 \big) \big(\mathcal{P}^{(+)}_2 + \mathcal{P}^{(-)}_2 \big) \nonumber \\
&=& - i \omega_{\bf k} \hat{P}_{i \sigma  {\bf k}} - i  \hat{P}^\dag_{m \bar \sigma  \bar{\bf k}}  \Bigg[ g^{++}_{\sigma \bk} \left( \mathcal{P}^{(+)}_1(t) \mathcal{P}^{(+)}_2(t)  \delta_{\sigma,+}+
\mathcal{P}^{(-)}_1(t) \mathcal{P}^{(-)}_2(t) \delta_{\sigma,-} \right)+ \nonumber \\
&& \hspace{3.5cm} g^{+-}_{\sigma \bk} \left( \mathcal{P}^{(-)}_1(t) \mathcal{P}^{(+)}_2(t) + \mathcal{P}^{(+)}_1(t) \mathcal{P}^{(-)}_2(t) \right) \delta_{\sigma,-\bar \sigma} \Bigg]\, .
\end{eqnarray}
%
The counter-circular term cancels out by destructive interference
\begin{eqnarray}\label{countercirc term}
&&\left( \mathcal{P}^{(-)}_1(t) \mathcal{P}^{(+)}_2(t) + \mathcal{P}^{(+)}_1(t) \mathcal{P}^{(-)}_2(t) \right) \delta_{\sigma,-\bar \sigma} \nonumber = \\
&& = \left( -i \mathcal{P}_{k_p}(t) \frac{e^{+i \theta}}{\sqrt{2}}
\mathcal{P}_{k_{p'}}(t) \frac{e^{-i \theta}}{\sqrt{2}} + i
\mathcal{P}_{k_p}(t) \frac{e^{-i \theta}}{\sqrt{2}} \mathcal{P}_{k_{p'}}(t)
\frac{e^{+i \theta}}{\sqrt{2}} \right) = 0\, ,
\end{eqnarray}
while the co-circular term reads
\begin{eqnarray}\label{cocirc term}
&& = \mathcal{P}^{(+)}_1(t) \mathcal{P}^{(+)}_2(t)  \delta_{\sigma,+}+
\mathcal{P}^{(-)}_1(t) \mathcal{P}^{(-)}_2(t) \delta_{\sigma,-} \nonumber = \\
&& = e^{-i \pi/2} \mathcal{P}_{k_{p}}(t) \mathcal{P}_{k_{p'}} (t)  \frac{e^{-i 2\theta}}{2} \delta_{\sigma,+} + e^{i \pi/2}\mathcal{P}_{k_{p}}(t) \mathcal{P}_{k_{p'}} (t) \frac{e^{+i 2\theta}}{2} \delta_{\sigma,-} = \\
&& =  \frac{e^{-i 2\theta}}{2}  e^{-i \pi/2} \left[  \mathcal{P}_{k_p}(t) \mathcal{P}_{k_{p'}} (t) \delta_{\sigma,+} - \mathcal{P}_{k_p}(t) \mathcal{P}_{k_{p'}} (t) e^{+i 4\theta} \delta_{\sigma,-} \right] \, . \nonumber 
\end{eqnarray}
Choosing the first pump such that $\theta=\pi/4$ and neglecting an overall phase, Eq.~\eqref{dtP_2pump_2} becomes 
\begin{eqnarray}\label{dtP_triplet}
\frac{d}{dt} \hat{P}_{i \sigma  {\bf k}} &=& - i \omega_{\bf k}
\hat{P}_{i \sigma  {\bf k}} - i G \hat{P}^\dag_{m \bar \sigma  \bar{\bf k}} \left( \delta_{\sigma,+} + \delta_{\sigma,-} \right)
\end{eqnarray}
with $G = g \mathcal{P}_{k_p}(t) \mathcal{P}_{k_{p'}} (t)$, and $g = g_{\sigma \bk} / 2$. This equation of motion can be obtained by the effective Hamiltonian Eq. (5) of the paper and ideally generates a pure entangled triplet state of the form

\begin{equation}
|\Psi\rangle = \frac{1}{\sqrt{2}} (|+,+\rangle +
|-,-\rangle)\, .
\end{equation}


\section{Acoustic phonon interaction}\label{sec:Phonons}

Long-wavelength acoustic phonon interaction Hamiltonian is obtained within the deformation potential coupling method. %
In general, moving from bulk systems to heterostructures, the partial loss of spatial symmetry leads to modifications not only in the electronic degrees of freedom but in the phonon subsystem as well.
However, detailed analysis of phonons in strainless 2D systems in III-V materials~\cite{SM_Rossi51PRB1995,SM_Rossi52PRB1995,SM_RuckerPRB1992} showed that exciton-phonon decoherence can be well described when assuming phonons as bulk-like.
The three-dimensional electron- (hole-) phonon interaction Hamiltonian for the deformation potential coupling has the form
\begin{equation}\label{H_int}
\hat H_{\text{int}} = \sum_{\veck \vecq} \left(\frac{\hbar q}{2 \rho u V} \right)^{1/2} \left(D_c \hat c^{\dag}_{\veck+\vecq} \hat c_{\veck} + D_v \hat d^{\dag}_{\veck+\vecq} \hat d_{\veck} \right) \left(\hat b_{\vecq} + \hat b^\dag_{-\vecq} \right), 
\end{equation}
where $V$ is the quantization volume, $u$ is the sound velocity, $\rho$ is the material density, $\veck$ is the 3D electron (hole) momentum, while $D_c$ and $D_v$ are deformation potential constants for the conduction and valence band, respectively. The operators  $\hat b_{\vecq}$  and $\hat b^\dag_{\vecq}$ are the bosonic annihilation and creation operators for  phonons of 3D momentum ${\vecq}$, with modulus $q$.

We want to derive the phonon-interaction Hamiltonian for quasi-two-dimensional 1S excitons in the case of GaAs/AlGaAs QW structures. This can be accomplished by projecting the full Hamiltonian in the exciton sector~\cite{SM_Takagahara}. In the case of a single QW, when considering only the lowest subbands in the confined direction, the exciton wave function depends only on the two-dimensional wave vector $\bkcm$ related to the center-of-mass motion and reads
\begin{eqnarray}\label{1Sexc}
\ket{1S; \bkcm} &=& \frac{a^3_0}{\sqrt{S}} \sum_{{\vec r}_e,{\vec r}_h} e^{-i \bkcm \cdot {\bf R}} W(\rho) \chi_e(z_e) \chi_h(z_h) \hat c^\dag_{c,{\vec r}_e} \hat c_{c,{\vec r}_h} \ket{0} \nonumber \\
&=& \sum_{\veck,\veck'} f(\bkcm;\veck,\veck') \delta_{\bk-\bk',\bkcm} \hat c^\dag_{\veck} \hat d^\dag_{- \veck'} \ket{0}.
\end{eqnarray}

The notation takes into account explicitly the separation between confined and free directions. Here a ``two-dimensional" exciton is written in terms of a three-dimensional envelope convolution (in direct or reciprocal space), $\veck = (\bk,k_z)$, ${\vec r} = ({\bx},z)$, $\rho = \bx_e - \bx_h$, $R = (m_e\bx_e+m_h \bx_h)/M$, $V=S L$, where $L$ is the confined direction quantization length and $S$ is the quantization surface in the free directions, capital letters are for the wave vectors of the center-of-mass of motion while the small $k$'s refer to the in-plane component of the three-dimensional wave vectors. 
The envelope function in reciprocal space reads ($\alpha_e = m_e/M, \ \alpha_h = m_h/M, \ M=m_e+m_h$)
 \begin{equation}
f(\bkcm;\veck,\veck') = \frac{1}{L^2} \int d^2 \rho \int dz_e \int dz_h W(\rho) \chi_e(z_e) \chi_h(z_h)  e^{-i (\alpha_e \bkcm - \bk) \cdot \rho} e^{-i k_z z_e} e^{i k'_z z_h}. 
\end{equation}
The reduction to the 1S exciton sector is obtained by an explicit projection:
\begin{equation}\label{projection}
\hat H^{\text{DF}}_\text{exc-ph} = \sum_{\bkcm \bkcm'} \bra{1S; \bkcm'} H_\text{int} \ket{1S; \bkcm} \ \ket{1S;\bkcm}\bra{1S;\bkcm'}.
\end{equation}
The procedure is fully reported in~\cite{SM_Takagahara}. The final results in the 1S exciton sector reads (we dropped the capital letter for brevity)
\begin{equation}\label{DF1well}
\hat H^{\text{DF}}_\text{exc-ph} = \sum_{\bk \bq} \sum_{q_z} \left(\frac{\hbar q}{2 \rho u V} \right)^{1/2} \left(D_c I^\bot_e(q_z) I^\|_e(\bq) + D_v I^\bot_h(q_z) I^\|_h(\bq) \right) \ket{\bk+\bq}\bra{\bk} \left(\hat b_{\bq,q_z} + \hat b^\dag_{-\bq,-q_z} \right).
\end{equation}
All the wave vectors in this Hamiltonian are now related to the center of mass of the exciton.
The two overlap integrals are given by
\begin{eqnarray}\label{overlap_integrals}
I^\bot_{e(h)}(q_z) &=& \int d z |\chi_{e(h)}(z) |^2 e^{i q_z z}, \\
I^\|_{e(h)}(\bq) &=& \int d^2 \rho |W(\rho)|^2 e^{i \alpha_{h(e)} \bq \cdot \rho} = \left(1+\left(\frac{m_{h(e)}}{2 M} q a_x\right)^2 \right)^{-3/2}.
\end{eqnarray}
The second integral in the plane is analytic owing to the form of the 1S exciton wavefucntion, $W(\rho) = \sqrt{\frac{2}{\pi a_x}} e ^{-\frac{\rho}{a_x}}$.

%

\subsection{Three quantum wells}

According to Eq.~(\ref{1Sexc}), the generic field operators for a 1S exciton can be expanded as ($\Phi^{\text(1S)}_{\bk}(\vecr) = \bra{\vecr} \hspace{-0.1cm} 1S,\bk \rangle$)
\begin{equation}\label{field_op_envelope}
\hat \Psi^\dag(\vecr) = \sum_{\veck,\veck'} \Phi^{\text(1S)}_{\bk}(\vecr) \hat c^\dag_{\veck_e} \hat d^\dag_{-\veck_h},\end{equation}
where $\Phi_{\bk}(\bx)$ is a shorthand notation for the $f$ above.
We are interested in the case of three quantum wells. We follow the spirit of Eq.~\eqref{field_op_envelope} and write ($\Phi^{\text(1S)}_{i,\bk}(\vecr) = \bra{\vecr} \hspace{-0.1cm} 1S;i,\bk \rangle$)
\begin{equation}\label{field_op_envelope2}
\hat \Psi^\dag(\vecr) = \sum_{n=1,2,3,\veck_e,\veck_h} \Phi^{\text(1S)}_{n,\bk}(\vecr) \hat c^\dag_{n,\veck_e} \hat d^\dag_{n,-\veck_h}.
\end{equation}
The non-diagonal matrix elements in Eq. (\ref{projection}) involve states belonging to different, space-separated, wells and the phonon interaction Hamiltonian will be zero because the overlap integrals over direct space equivalent to Eq. (\ref{overlap_integrals}) are zero. 
Taking as the origin in z the well number 2, the matrix element
\begin{equation}
\bra{2; \bk'} \hat H_\text{int} \ket{2; \bk},
\end{equation}
has already been calculated as in the single well case, whereas the corresponding matrix element for QW 1 and 3 are basically identical except for the overlap integral in the z direction which has to be evaluated over the corresponding well (centered around $\pm\text{z}_\text{qw}$ and of the same width as the other one). 
%
Eventually,  by introducing 
\begin{equation}\label{DF1well}
\Xi(\bq,q_z) = \left(\frac{\hbar q}{2 \rho u V} \right)^{1/2} \left(D_c I^\bot_e(q_z) I^\|_e(\bq) + D_v I^\bot_h(q_z) I^\|_h(\bq) \right)\, ,
\end{equation}
the Hamiltonian for the triple cavity structure becomes

\begin{eqnarray}\label{SM_DF3well}
\hat H^{\text{DF}}_\text{exc-ph} && = \sum_{\bk,\bq,q_z,n=-1,0,1} \Xi(\bq,q_z) \ket{n ;\bk+\bq}\bra{n; \bk} \left(e^{i q_z \lambda_n} \hat b_{\bq,q_z} + e^{-i q_z \lambda_n} \hat b^\dag_{- \bq,-q_z} \right),
\end{eqnarray}
%
where here we have numbered the wells as $n=-1,0,1$, and $\lambda_n= n \text{z}_\text{qw}$.
From this Hamiltonian we calculate the scattering rates $W_{\bf k,\bf k'}$ for the Boltzmann equation as outlined in Ref~\cite{SM_PortolanPRB2008_Langevin}. The numerical values for the deformation potential constants are taken from~\cite{SM_PiermarocchiPRB1996}.


\section{Numerical simulation of the Boltzmann equation}\label{sec:Numerics}

Using Fermi's golden rule, the linear PL dynamics for the populations $N^{\text{PL}}$ can be described microscopically by a Boltzmann equation
\begin{equation}\label{SM_Boltzmann}
\partial_t N^{\text{PL}}_{i \bk} =  -\Gamma_{i \bk} N^{\text{PL}}_{i \bk} + I_{i \bk} + \sum_{l \bk'} W^{(\text{ph})}_{i \bk,l \bk'} N^{\text{PL}}_{l \bk'},
\end{equation}
where the total linewidth $\Gamma_{i \bk} = \Gamma^{(\text{ph})}_{i \bk} + \gamma^{\text{(rad)}}_{i \bk}$ includes phonon-induced and radiative losses, which are independent of spin, and the scattering rates $W^{(\text{ph})}_{i \bk,l \bk'}$ are defined in the main text.
%
For simulating the external driving field at a specific $\bk$ we take a Gaussian pulse centered around a wave vector $\bk_p$,
\begin{equation}\label{1D_pump1}
I_{i \bk} = I^0_{2D} \frac{e^{-\frac{|\bk-\bk_p|^2}{2 \sigma^2}}}{2 \pi \sigma^2}\,.
\end{equation}
%
%
The resulting coherent part of the population is then given by
\begin{equation}\label{eq:N_coh}
\partial_t N^{\text{PL}}_{i \bk}\big|_{\text{coh}} =  -\Gamma_{i \bk} N^{\text{PL}}_{i \bk}\big|_{\text{coh}} + I_{i \bk}.
\end{equation}
In the paper we are mainly interested in the (incoherent) background photoluminescence defined as $\tilde N^{\text{PL}}_{i \bk} = N^{\text{PL}}_{i \bk} - N^{\text{PL}}_{i \bk} \big|_{\text{coh}}$. 
%
In general, the background photoluminescence can be calculated numerically by solving Eq.~\eqref{SM_Boltzmann} and Eq.~\eqref{eq:N_coh} on a 2D grid. 


\subsection{Equivalent symmetric problem}

In experimental and numerical studies it is found that the incoherent steady-state polariton population $\tilde N^{\text{PL}}_{i \bk}$ has radial symmetry even under excitation with a specific wave vector \cite{SM_LangbeinPRB2004,SM_PortolanPSSc2008}. 
This property can be seen as peculiar of the quasi-elastic nature of the phonon-induced scattering $W^{(\text{ph})}$ in this system, which is able to redistribute in a very short time the peaked coherent population in a symmetric pattern and then the steady-state incoherent population depends only on the total flux of injected particles balanced by loss and no more on the specific form of $N^{\text{PL}}_{i \bk}\big|_{\text{coh}}$ at earlier times.

In our numerical studies this allows us to approximate the full 2D problem by an equivalent one with radial symmetry from the outset. The population calculated in this way should well reproduce the full 2D incoherent population once the steady-state is reached. Moreover, the symmetry of this equivalent system reduces the computation to the sole radial distribution for the population $N^{\text{PL}}_{i k}$, where $k=|\bk|$. 

To do so we consider a radially symmetric pump 
\begin{equation}\label{1D_pump1}
\bar I_{i \bk} = I^0 \frac{e^{-\frac{(k-k_p)^2}{2 \sigma^2}}}{2 \pi \sigma^2},
\end{equation}
which is chosen such that the flux of injected particles into the system is the same as in the case of a single pumped wave vector,
\begin{equation}\label{flux_condition}
\sum_{k_x,k_y} I_{i \bk} = \sum_{k_x,k_y} \bar I_{i \bk}.
\end{equation}
We change to polar coordinates,
\begin{equation}
\sum_{k_x,k_y} \simeq \frac{S}{(2 \pi)^2} \int dk_x dk_y =  \frac{S}{(2 \pi)^2} \int k dk d\theta \simeq \sum_{k,\theta} \left(\frac{S}{(2 \pi)^2} \Delta k \Delta \theta \right),
\end{equation}
and by making use of the fact that the distribution of populations is radially symmetric, $N^{\rm PL}_{i k\theta}\equiv N_{ik}^{\rm PL}$, we can manipulate the last term in Eq.~\eqref{SM_Boltzmann} to obtain
\begin{equation}\label{2D_eq_2}
\partial_t N^{\rm PL}_{i k} = - \Gamma_{i k}  N^{\rm PL}_{i k} + \bar I_{i k} + \sum_{l k'} W_{(ik), (lk')} N^{\rm PL}_{l k'}\, ,
\end{equation}
where we introduced $W_{k, k'}= \sum_{\theta'} W_{(k \theta), (k' \theta')}$.  In Fig. (2) of the main text we solve the stationary solution of this equation and plot the stationary values of the incoherent part $\tilde N^{\rm PL}_{i k}$.

\vspace{1.0cm}

\end{widetext}


\begin{thebibliography}{50}
\bibitem{g_specification}
In particular, for signal-idler wavevector $\bk,\bk_i$, the pump-mixed process in Fig. \ref{Coupled_cavities} has a nonlinear coefficient of the form $ g = \frac{1}{2} \sum_p\left( V/n_{\text{sat}} \right) X^p_{2 \bk_s} X^p_{2 \bk_i}\left( X^p_{3 \bk_{p1}} X^p_{1 \bk_{p2}} \right.$ $ + \left. X^p_{3 \bk_{p2}} X^p_{1 \bk_{p1}} \right) + 2 V_{\text{xx}} X^p_{2 \bk_s} X^p_{2 \bk_i} X^p_{2 \bk_{p1}} X^p_{2 \bk_{p2}} $, 
with the saturation density $n_{\text{sat}}$ and the Coulomb-induced contribution including its correction beyond mean-field $V_{\text{xx}}$.
\end{thebibliography}
\end{document}